\documentclass[reprint,11pt,authaffil]{JASAnew}
\usepackage[T1]{fontenc}
\usepackage[utf8]{inputenc}
\usepackage{amsmath,amssymb}
\usepackage{graphicx}
\usepackage{booktabs}
\usepackage{array}
\usepackage{multirow}
\usepackage{tabularx}
\usepackage{makecell}
\usepackage{siunitx}
\usepackage{xcolor}
\usepackage{url}
\usepackage{hyperref}
\usepackage{enumitem}
\usepackage{placeins}
\usepackage{bm}
\usepackage{ragged2e}

\graphicspath{{figures/}}
\setcitestyle{authoryear,round,semicolon}
\hypersetup{
  hidelinks,
  pdftitle={Structure Across Voices: Comparing acoustic-event type accumulation and sequence dependence across four vocal corpora using frozen audio encoders},
  pdfauthor={Mudit Sinha and Sanika Chavan},
  pdfkeywords={animal bioacoustics, cross-species acoustics, acoustic-event type accumulation, sequence dependence, sperm whale, birdsong, primate vocalization, frozen audio encoders}
}
\renewcommand{\arraystretch}{1.12}

\makeatletter
\def\enoteheading{}

\makeatother
\ifmanuscript\fi

\ifmanuscript\else
\fi

\begin{document}
\title[Structure Across Voices]{Structure Across Voices: Comparing acoustic-event type accumulation and sequence dependence across four vocal corpora using frozen audio encoders}

\author{Mudit Sinha}
\email{muditsinha01@gmail.com}
\affiliation{Independent Researcher, San Bruno, California 94066, United States}
\author{Sanika Chavan}
\email{sanikac10@gmail.com}
\affiliation{Independent Researcher, San Bruno, California 94066, United States}

\date{16 November 2026}

\begin{abstract}
Comparing acoustic structure across very different vocal corpora is difficult because the recordings use different native event units, contain different amounts of sequence, and were collected under different conditions. We use the high-dimensional representations of frozen, pretrained audio encoders to measure acoustic diversity and temporal organization. Matched sampling gives each corpus the same opportunities to observe new acoustic forms and relationships between events, without changing its native event definition. We compare sperm-whale codas, human speech phones, Bengalese-finch syllables, and common-marmoset calls using 1,952 events and 1,700 adjacent pairs per comparison. Whale shows the fastest accumulation of encoder-defined acoustic groups, while Finch has the strongest repeated-sequence recurrence against ordinary within-block shuffles. Their neighboring-event dependence is closely matched under a source-and-position control. A second preceding event changes held-out prediction more favorably for Whale than for the other corpora, although Whale's standalone improvement remains uncertain. Waveform measurements recover several differences, but duration and acquisition-sensitive properties also contribute. Together, the findings show how pretrained representations can measure distinct forms of vocal organization, while waveform and source controls define the scope of the comparison.
\end{abstract}
\maketitle
\begin{figure*}[!t]
\centering
\includegraphics[width=0.97\textwidth]{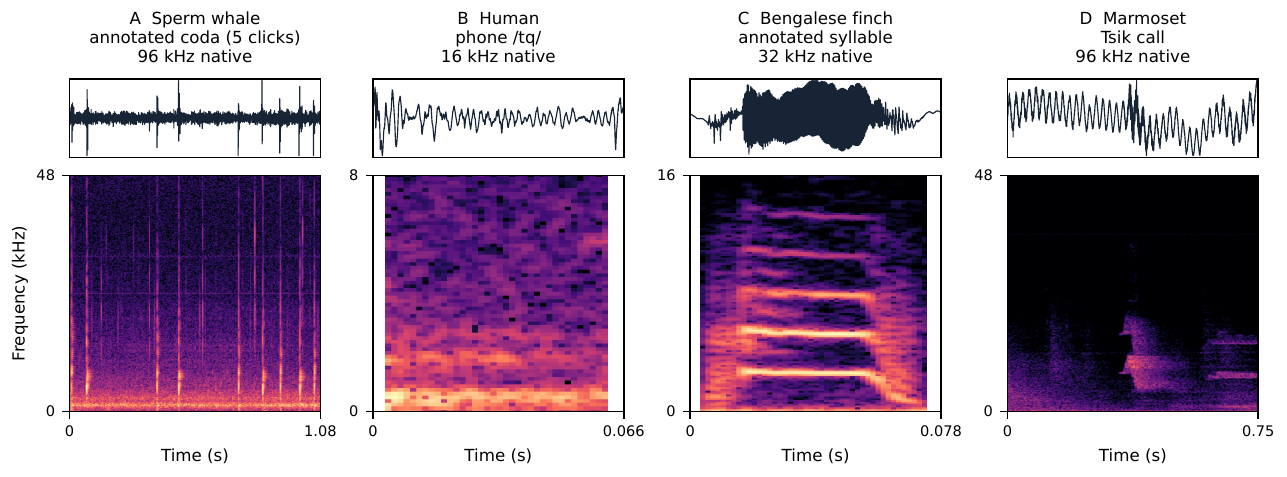}
\caption{Example native events in the matched corpora: a sperm-whale coda, a human phone, a Bengalese-finch syllable, and a common-marmoset call. Each panel retains its own time and frequency range; the Whale inputs include remote 96-kHz recordings and 60-kHz tag recordings. Waveform and spectral displays are normalized within each event. The displayed Whale click count is an annotation count, not an independently validated coda type.}
\label{fig:native-events}
\end{figure*}

\section{Introduction}

Very different vocal corpora can be placed side by side, but deciding what constitutes a fair acoustic comparison is difficult. A sperm-whale coda, a human speech phone, a Bengalese-finch syllable, and a common-marmoset call are different kinds of events, recorded and annotated under different conditions. They can also contain very different amounts of data and sequence. A corpus with more recorded events has more chances to reveal new acoustic forms, transitions, and repeated patterns. The challenge is to compare shared properties of these recordings without first forcing their sounds into the same biological or temporal unit. Figure~\ref{fig:native-events} shows how different the native events considered here can be.

We address this problem with two complementary forms of standardization. Several pretrained audio encoders, held fixed throughout this study, describe each recorded event through many learned acoustic features. Rather than using these representations primarily to group similar sounds, we use them to investigate acoustic variation and the organization of sounds over time. Acoustic groups are formed separately within each corpus when a measurement requires them. Repeating the comparison across encoder choices tests whether the result depends on one particular representation. We then match the amount of evidence available to each corpus, giving them the same number of events and the same opportunities to observe neighboring and longer sequences. The native event definitions are preserved, while recording and acquisition differences remain part of the interpretation.

With a common acoustic description and matched observation opportunity, we can ask how these very different corpora differ in practice. We compare the appearance of new acoustic forms, the organization of neighboring sounds, and the recurrence of longer patterns. We then ask whether an additional earlier event helps predict what comes next. Finally, we examine the same recordings through direct waveform measurements and controls that preserve recording source or sequence position. These questions address different aspects of the sounds and their order. Together they give a more useful comparison than a single measure of vocal complexity.

The comparison reveals different acoustic and temporal profiles. Whale recordings show the fastest accumulation of new acoustic groups, while Finch recordings show especially strong recurrence of repeated group patterns. The comparison changes when we preserve source and sequence position in the order control: neighboring-event dependence in Whale and Finch is then closely matched. Direct measurements of the recorded waveforms recover parts of the pattern, but also reveal the influence of duration and recording conditions. The result is a comparison of distinct measurable properties, rather than a claim that one vocal system is generally more complex. We can identify where the recordings differ, how strongly, and under which acoustic and temporal measurements those differences become visible.

\section{Related Work}
Comparative bioacoustic studies measure differences in acoustic form,
repertoire composition, and sequence organization across species and
datasets \citep{kershenbaum2016review,odom2021comparative}. A difficulty
is that calls, syllables, codas, and speech phones are defined at different
scales and can differ substantially in duration and acoustic structure.
The choice of event unit therefore shapes the comparison itself. We retain
each corpus's native event definition and compare the same acoustic and
temporal properties without changing the units being studied.

A second challenge is describing acoustically different events through a
common procedure. \citet{sainburg2020latent} used low-dimensional acoustic spaces
constructed from vocalization spectrograms to study repertoires and
sequence structure, while \citet{morfi2021deep} studied animal-sound
similarity using learned representations. \citet{best2023deep}
used deep audio embeddings, including pretrained models, to cluster
vocalizations and compare the groups with expert annotations.

These studies show that learned representations can reveal acoustic structure.
Here, we use the high-dimensional representations of pretrained, frozen audio encoders as spaces in which to measure that structure. The encoders have already learned many acoustic features before we apply them to the present corpora. We use those features to investigate how sounds vary and how their relationships change over time. Acoustic groups are introduced within each corpus only when a statistic needs discrete categories. Other analyses measure continuous acoustic variation without forming groups. The question is therefore what the representations reveal about recorded vocal organization, rather than how well a particular clustering recovers expert labels.

Sequence estimates also depend on the amount and continuity of available
data \citep{mccowan1999quantitative,suzuki2006entropy,sainburg2019parallels}.
A larger or less fragmented corpus has more opportunities to reveal acoustic
forms, transitions, and repeated patterns. We therefore match both the
number of events and the local sequence opportunities before making
cross-corpus comparisons. This helps distinguish differences in measured
structure from differences in how much sequence evidence is available.
Equal observation opportunity does not, by itself, establish equal acoustic
resolution across the different native event scales. We therefore test the
patterns with direct waveform measurements, analyses without clustering, and
source-aware randomizations.

\section{Methods}
We compare four vocal corpora using the same questions while preserving each corpus's native events and the recording boundaries supported by its metadata. The procedure has three parts. First, a frozen audio encoder describes the acoustic properties of each recorded event. Second, we match the amount of evidence available to every corpus, including the number of events and the opportunities to observe sequences. Third, we measure how acoustic forms accumulate and how the order of events is organized. Additional waveform and source-aware analyses test which parts of the result remain visible under other acoustic descriptions and more restrictive controls. Table~\ref{tab:overview} shows what remains specific to each corpus and what is held common.
\begin{table*}[!t]
\caption{Comparison design. Equal observation opportunity does not imply equal event meaning or physical acoustic resolution.}
\label{tab:overview}
\centering\small
\setlength{\tabcolsep}{3pt}
\renewcommand{\arraystretch}{1.18}
\begin{tabular}{@{}lll@{}}
\toprule
\parbox[t]{.17\textwidth}{\textbf{Step}} & \parbox[t]{.29\textwidth}{\textbf{Retained from the corpus}} & \parbox[t]{.41\textwidth}{\textbf{Held common across the comparison}} \\
\midrule
\parbox[t]{.17\textwidth}{Native event} & \parbox[t]{.29\textwidth}{Whale coda, Human phone, Finch syllable, Marmoset call} & \parbox[t]{.41\textwidth}{One annotated event per measurement} \\
\parbox[t]{.17\textwidth}{Source and chronology} & \parbox[t]{.29\textwidth}{Within-file onset order and source boundaries} & \parbox[t]{.41\textwidth}{No transitions across unsupported boundaries} \\
\parbox[t]{.17\textwidth}{Acoustic description} & \parbox[t]{.29\textwidth}{Corpus-specific distributions of encoder vectors} & \parbox[t]{.41\textwidth}{Same frozen encoder procedure in each configuration} \\
\parbox[t]{.17\textwidth}{Observation budget} & \parbox[t]{.29\textwidth}{63 Whale recordings; four conservative source pools} & \parbox[t]{.41\textwidth}{1,952 events; 252 maximum-16-event blocks; 1,700 adjacent pairs} \\
\parbox[t]{.17\textwidth}{Measurement} & \parbox[t]{.29\textwidth}{Native acoustic and sequential variation} & \parbox[t]{.41\textwidth}{Three primary statistics and specified order nulls} \\
\parbox[t]{.17\textwidth}{Checks} & \parbox[t]{.29\textwidth}{Recording, source, duration, and event-position differences} & \parbox[t]{.41\textwidth}{Waveform, continuous, padding, and source-aware analyses} \\
\bottomrule
\end{tabular}
\end{table*}

\subsection{Event units}

Each corpus provides its own annotated vocal units, and we retain those
units rather than imposing a common segmentation across species.

For sperm whales, one event is a \textit{coda}: a short sequence of
broadband clicks whose temporal pattern is commonly described by the
intervals between successive clicks \citep{watkins1977,sharma2024}.

For the Human corpus, one event is an annotated \textit{phone}: one realized
speech sound, such as a consonant or vowel segment, identified within
continuous conversational speech \citep{pitt2007}.

For Bengalese finches, one event is an annotated \textit{syllable}: one
bounded vocal element within a song \citep{koumura2016}.

For common marmosets, one event is an annotated \textit{call}: one bounded vocalization identified from its onset to its offset in the recording
\citep{lamothe2025}.

These units differ in biological meaning and scale and are treated only as the native observation units provided by each corpus.

\subsection{Datasets and matched sequence evidence}
We selected four annotated corpora because they span different event scales and recording conditions while preserving enough event order for the questions tested here. The Human corpus contains conversational phones from Buckeye~\citep{pitt2007}. The Finch corpus contains annotated Bengalese-finch syllables~\citep{koumura2016,koumura2016dataset}, and the Marmoset corpus contains bounded vocal calls from MarmAudio~\citep{lamothe2025}. We refer to the four corpora as Whale, Human, Finch, and Marmoset. These names describe datasets, not pooled estimates for whole species.

The Whale corpus contains 1,952 annotated coda crops from 63 recordings in the publicly shared SpermWhales collection~\citep{gubnitsky2024clicks,gubnitskyDiamantPublicSpermWhales}: 33 remote WAV files at 96 kHz and 30 tag FLAC files at 60 kHz. The derived representations come from \texttt{ceti-shared-embeddings-v1} (revision \texttt{710974f}). A supplementary manifest identifies each recording--annotation pair and its acquired-file checksum. The metadata do not establish individual whale identity or social-unit labels for every event. The Supplement documents crop matching and the operational source pools.

Events retain annotated onset order within each recording. Coincident onsets, annotation aliases, gaps, and file boundaries are handled before any sequential measurement. Accepted runs are divided into operational blocks containing at most 16 events. The final Whale pattern contains 252 blocks and 1,952 events. It supplies 1,700 adjacent pairs and 1,491 positions with two predecessors; the 1,307 positions shared by order-0 through order-3 predictors form a further restricted held-out target set. These numbers describe available statistical opportunities, not verified uninterrupted behavioral bouts.

Matching event count alone would not give the corpora equal sequence evidence. A corpus with the same number of events concentrated in a few long sequences has more chances to exhibit transitions and recurring patterns than one split among many short sequences. We therefore reproduce the accepted Whale block-length pattern within each reference corpus using contiguous eligible events from its native sources. Fifty matched draws from each reference pool vary which observations are chosen. Every draw has the same event budget and opportunities for adjacent and longer order comparisons. The 50 draws are alternative selections, not 50 independent animals or recording sources.

Recording sources also determine which events can legitimately contribute independent evidence. For Whale, four conservative recording-source pools cover the 63 files: one pool with unresolved remote-recording identity and three labeled recording pools. These pools are not four verified animals or social units. Human uses recording tracks, Finch uses birds, and Marmoset uses available recording-day groups. Source-bootstrap inference resamples these operational sources after averaging over analytical draws. The supplement reports the exact shell, exclusions, and source-pool counts.

\subsection{Frozen acoustic representations}
The four corpora do not share acoustic category labels, so we need a consistent way to describe their sounds before measuring them. Each event waveform is passed through an unchanged pretrained audio encoder, which converts the waveform into numerical acoustic features learned previously from audio. Its frame-level outputs are pooled into one vector per event. These vectors describe acoustic similarity continuously; they are the measurements used before any discrete groups are introduced.

We use 11 configurations spanning HuBERT, wav2vec~2.0, AVES, AVES2, OpenBEATs, and VampNet~\citep{hsu2021hubert,baevski2020wav2vec,hagiwara2023aves,miron2026avex,bharadwaj2025openbeats,flores2023vampnet}. These encoders were developed for different audio domains, including speech and broader sounds. Within a configuration the same preparation and pooling procedure is applied to every corpus. Using different pretrained models tests dependence on one learned description of sound. Exact checkpoint provenance and any possible training-data overlap require confirmation from the original manifests; the fixed weights used for analysis alone do not establish that the training data are completely independent of the recorded corpora.

Some events are shorter than the 0.175-s minimum encoder input. These are extended before feature extraction. Both right-zero and symmetric-reflect padding are evaluated, along with level normalization. The additional signal can influence encoder output and group assignment, although it does not change the order of events. We distinguish input-duration effects from order effects in the sensitivity analysis.

Our main categorical sequence measurements require recurring acoustic groups. We use K-means~\citep{macqueen1967methods}: for a chosen $K$, the algorithm places centers in the feature space and assigns each event to its nearest center. The centers are updated from the event vectors until the partition stabilizes. We do not use the annotated phone, syllable, call, or coda labels to define these groups.

We fit K-means separately within each corpus and encoder configuration at $K\in\{16,24,32\}$. This allows acoustic groups to follow each corpus's recorded variation, rather than forcing the same physical centers on unlike sounds. Group 7 in Whale has no assumed correspondence to group 7 in Finch. Equal $K$ makes the number of categories available to the analysis equal; it does not make their acoustic distances or biological meaning equal. In the descriptive comparison, 11 configurations, 50 draws, and three group counts give 1,650 analytical cells. These are settings, not independent biological replicates.

For the more restricted source-and-position comparisons, a fixed family-balanced consensus defines $K=32$ groups. The longer-history held-out analysis instead relearns its consensus and preprocessing on training sources only. The fixed and train-only group constructions answer different questions, so their estimates are not treated as interchangeable.

\subsection{Analysis}
We study two different properties of the recordings: \textit{acoustic diversity} and \textit{temporal organization}. Acoustic diversity asks how broadly the events vary as more examples are observed. Temporal organization asks whether the acoustic properties of one event are related to those of events nearby in the sequence. A corpus can contain many distinguishable sounds without placing them in a strongly repeated order. Another corpus can use fewer distinguishable sounds but arrange them in more predictable sequences.

We therefore examine temporal organization at several scales. One measurement tests the relationship between immediate neighbors, another asks whether longer patterns recur, and a held-out prediction test asks whether a second preceding event adds useful information. These measurements use the same event boundaries and matched sequence opportunities, but they describe different properties. The controls below help determine how much depends on acoustic grouping, source identity, and event position.

\subsection{Measurements}
\medskip
\noindent\textit{(1) Type accumulation (acoustic diversity).}
We ask how quickly previously unseen acoustic groups continue to appear as more events are observed. If $V(n)$ is the number of distinct groups encountered after $n$ events, the Heaps--Herdan exponent $\beta$ summarizes the growth~\citep{herdan1964quantitative,heaps1978information}:
\[
V(n)\propto n^{\beta}.
\]
A larger $\beta$ means faster accumulation within the matched event range and the chosen group resolution. Because every corpus uses the same number of events and available groups, the procedure controls the opportunity to encounter new categories. The exponent describes the finite range of events examined here. It does not count the natural vocal types of a species or measure the ultimate size of its repertoire. It tells us how quickly the available acoustic groups become represented under the same sampling conditions.

\medskip
\noindent\textit{(2) Neighboring-event dependence (temporal organization).}
We next ask whether one event's group reduces uncertainty about the group of the next event. We estimate this reduction using a Miller--Madow-corrected categorical entropy calculation~\citep{miller1955bias}. To ask what comes from the observed order, we subtract the mean dependence after randomizing group order within each block. This shuffle keeps the events, group frequencies, and block lengths but breaks their original neighborhood relationships. Positive excess, reported in bits per event, means the actual order contains more information about the next group than these shuffled sequences.

A shuffled sequence is useful for asking whether the recorded order matters, but it does not eliminate every alternative explanation. Sounds from the same source can resemble one another because they were recorded under similar conditions. Some groups may also occur preferentially near the beginning or end of a block. Both effects can make neighboring events appear related. We therefore perform a second test that preserves the source pool and eight bins of normalized block position. This restricted comparison uses a fixed, family-balanced consensus at $K=32$. Because both the group construction and the randomized baseline differ from the ordinary shuffle, the two excess values answer different questions and should not be directly subtracted.

\medskip
\noindent\textit{(3) Repeated-sequence recurrence (temporal organization).}
Neighboring dependence concerns only adjacent events and does not show whether longer patterns recur. We therefore test whether repeated multi-event strings make the observed group sequences easier to compress than randomized sequences. RePair replaces common symbol pairs with reusable rules. LZ78 builds a dictionary of previously encountered phrases, and DEFLATE uses string matching and entropy coding~\citep{larsson2000repair,ziv1978compression,deutsch1996deflate}. The recurrence implementation uses the LZ78 dictionary algorithm.

For each procedure, we divide the compressed size of the observed sequences by the average compressed size after randomizing order within the same blocks. A ratio below one means that the observed sequence contains more reusable patterns than the same events arranged in shuffled order. We also calculate ratios against the more restrictive source-and-position null. Its ratios are reported separately because that control preserves relationships that the ordinary shuffle removes. The three compression procedures detect different kinds of repetition. We therefore compare their results rather than averaging them into one measure of temporal structure.

\medskip
\noindent\textit{Secondary analysis: Longer temporal context.}
To examine history beyond one adjacent event, we compare held-out prediction using one or two earlier groups. In a sequence A--B--C, order~1 predicts C using B, and order~2 additionally uses A. The reported difference is order-2 minus order-1 held-out log likelihood, in bits per event. A positive difference means that the earlier event provides additional predictive information on the common held-out targets.

For this test, acoustic features, principal components, group centers, and consensus construction are fitted using training sources only. A family-balanced 11-configuration panel and $K=32$ support the primary analysis, with the same 1,307 target positions for orders 0--3 in each draw. The 50 draws are combined within source before source-bootstrap inference. With only four conservative Whale pools, an exact two-sided source sign-flip test has a minimum nonzero $p$ value of 0.125. The separately planned Whale-versus-reference contrasts use simultaneous intervals and Holm adjustments~\citep{holm1979simple}. These different uncertainty calculations answer different questions.

\subsection{Controls}
The primary measures use learned representations and, for three analyses, discrete groups. Recording source or event position could also contribute to apparent structure. We therefore use independent acoustic measurements and controls designed to distinguish what each observation establishes.

\medskip
\noindent\textit{Direct waveform measurements.}
We measure familiar physical properties of the original recordings: event duration and envelope timing, spectral location and extent, and spectral shape. We also examine acquisition-sensitive measures, including level and channel-related properties, as potential explanations rather than assuming they contain biological structure. Spectral feature families are evaluated within a common 0--8-kHz band. Each family is processed on the same matched shell, and its three-measurement profile is compared descriptively with the encoder-based profile.

\medskip
\noindent\textit{Continuous acoustic analyses.}
Our type and order measurements depend on group labels, so we also examine continuous physical acoustic descriptions without K-means. Coverage tests how quickly additional events fill the measured acoustic space; next-event prediction asks whether the physical properties of one event help predict the next. These test related forms of diversity and dependence but do not estimate exactly the same statistic as a categorical group analysis.

\medskip
\noindent\textit{Source and sequence position.}
Events from the same recording source may share conditions or acoustic properties. We therefore distinguish between-source variation from patterns that remain within sources. A separate position-conditioned randomization preserves source and location within a block. This allows a stronger question than the ordinary shuffle: whether order-specific structure remains after conditioning on where in each source sequence an event occurs.

\medskip
\noindent\textit{Analysis choices.}
We repeat the comparisons across encoder and clustering settings, alternative matched selections, reflect padding, normalization, and source-aware uncertainties. For longer-history sensitivity, we compare a 23-view representation panel (Literature23) with the 11-configuration family-balanced panel (Ladder11). Full estimates, accepted input checks, and numerical provenance are reported in the Supplementary Material. These analyses assess whether the results depend on a particular measurement procedure; they do not by themselves equate the physical sensitivity of different encoders.
\section{Results}
The four corpora differ in the ways their sounds vary and recur. Whale has the fastest accumulation of acoustic groups, while Finch shows the clearest repeated patterns under ordinary within-block shuffling. Their local-order comparison becomes less distinct after source and event position are preserved. We first examine these three measurements together, and then ask what changes when prediction uses two preceding events instead of one. Finally, we turn to waveform measurements and source-aware controls to test which patterns are supported by the recorded sounds and which depend on the measurement procedure.

\begin{figure*}[!t]
\centering
\includegraphics[width=0.985\textwidth]{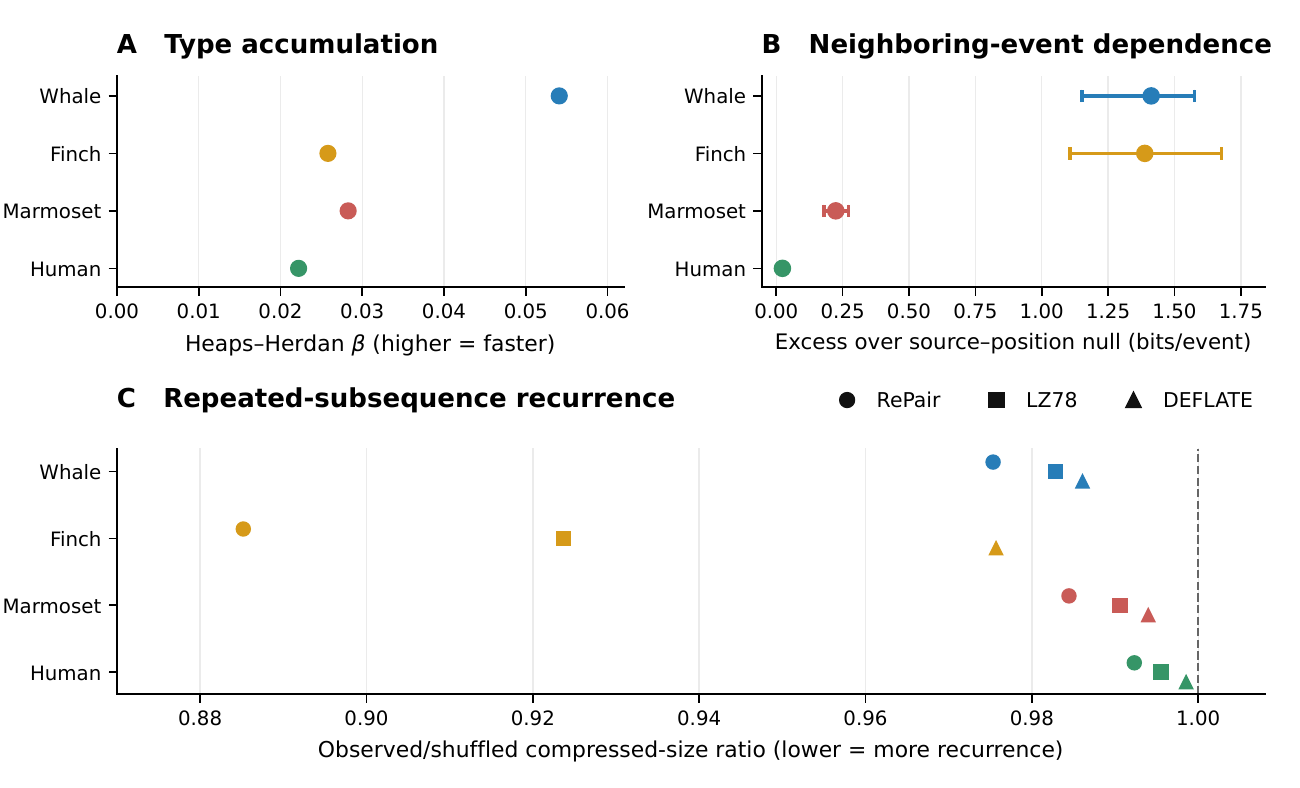}
\caption{Three complementary measurements on the 1,952-event matched shell. (A) Mean Heaps--Herdan exponent across 1,650 analytical cells; larger values indicate faster accumulation of encoder-defined groups. (B) Neighboring-event excess for the family-balanced fixed $K=32$ consensus relative to a source-and-position null; bars show 95\% source-bootstrap intervals. (C) RePair, LZ78, and DEFLATE observed-to-shuffled recurrence ratios under ordinary within-block randomization; smaller ratios indicate more repeated ordering. Panels B and C use different nulls and grouping scopes.}
\label{fig:profile}
\end{figure*}

\subsection{The four corpora show different acoustic and temporal profiles}
Type accumulation gives the clearest difference in acoustic diversity (Fig.~\ref{fig:profile}A). Whale has the highest mean Heaps--Herdan exponent at 0.0541, followed by Marmoset at 0.0283, Finch at 0.0258, and Human at 0.0222. This exponent describes how quickly previously unseen acoustic groups appear as more events are included. Every corpus has the same event budget and number of available groups in a given comparison. New groups therefore continue to appear most rapidly in Whale under those shared observation conditions. The result concerns the acoustic groups defined by the encoders; it does not estimate the total number of natural coda types.

Finch stands out most clearly when we examine recurring patterns of group labels. All three compression procedures give Finch the smallest observed-to-shuffled ratio under the ordinary within-block null (Table~\ref{tab:recurrence-ratios}; Fig.~\ref{fig:profile}C). RePair gives 0.885 for Finch and 0.975 for Whale. LZ78 gives 0.924 and 0.983, while DEFLATE gives 0.976 and 0.986. Because smaller ratios indicate greater compression relative to randomized order, these values show that Finch's observed groups recur more systematically than expected from the same groups in shuffled order. The result concerns patterns of analysis-defined acoustic groups, rather than a demonstrated syntax or communication function.

\begin{table}[!htbp]
\caption{Repeated-sequence recurrence relative to ordinary within-block shuffling. Lower ratios mean the observed order is more compressible than shuffled order. Means are over the 1,650 descriptive analysis cells; the source-and-position restricted-null ratios are reported separately.}
\label{tab:recurrence-ratios}
\centering\scriptsize
\setlength{\tabcolsep}{3pt}
\begin{tabular}{@{}lrrrr@{}}
\toprule
Method & Whale & Finch & Human & Marmoset \\
\midrule
RePair & 0.9753 & \textbf{0.8852} & 0.9923 & 0.9845 \\
LZ78 & 0.9828 & \textbf{0.9237} & 0.9955 & 0.9906 \\
DEFLATE & 0.9861 & \textbf{0.9757} & 0.9986 & 0.9940 \\
\bottomrule
\end{tabular}
\end{table}

Neighboring-event dependence asks a different question: does knowing the acoustic group of one event tell us something about the next event? Against the ordinary within-block shuffle, Finch shows the largest excess at 0.698 bits/event, followed by Whale at 0.136, Marmoset at 0.117, and Human at 0.065. The shuffle keeps the same events in each block but changes their order. These excess values therefore describe the information carried by the observed group-to-group arrangement beyond that random-order reference. They do not measure how much acoustic information is contained within an individual waveform.

The comparison changes when the control retains more of the source structure. At the fixed family-balanced $K=32$ consensus, a randomization that preserves source identity and relative block position gives 1.388 bits/event for Finch and 1.412 for Whale (Fig.~\ref{fig:profile}B). The Finch-minus-Whale difference is $-0.024$ bits/event, with a 95\% source-bootstrap interval of $[-0.379,0.418]$. We cannot distinguish the two corpora on this restricted comparison.

This does not contradict the ordinary-shuffle result. The first comparison removes most event ordering within blocks. The second retains source and position relationships that could otherwise contribute to apparent dependence. It also uses a different grouping construction. The change tells us that Finch's lead is sensitive to what the null model preserves, rather than showing that one of the estimates is incorrect.

A descriptive summary that assigns equal weight to type accumulation, shuffle-adjusted local dependence, and compression recurrence gives Finch a mean rank of 1.393 and Whale 1.863 across the analytical settings. The summary is not a biological complexity score. The separate measurements show why: Whale leads one kind of acoustic diversity, Finch leads ordinary-shuffle recurrence, and the local-order comparison depends on what structure is held fixed.

\subsection{Using two previous events instead of one shifts the temporal comparison toward Whale}
We next asked whether information beyond the immediately preceding event helps predict the next event. For A--B--C, the order-1 model predicts C from B, whereas the order-2 model also uses A. All models are evaluated at identical held-out targets, and group construction uses training sources only. This avoids allowing the held-out recordings to define the groups used to score their own sequences.

\begin{figure*}[!t]
\centering
\includegraphics[width=0.92\textwidth]{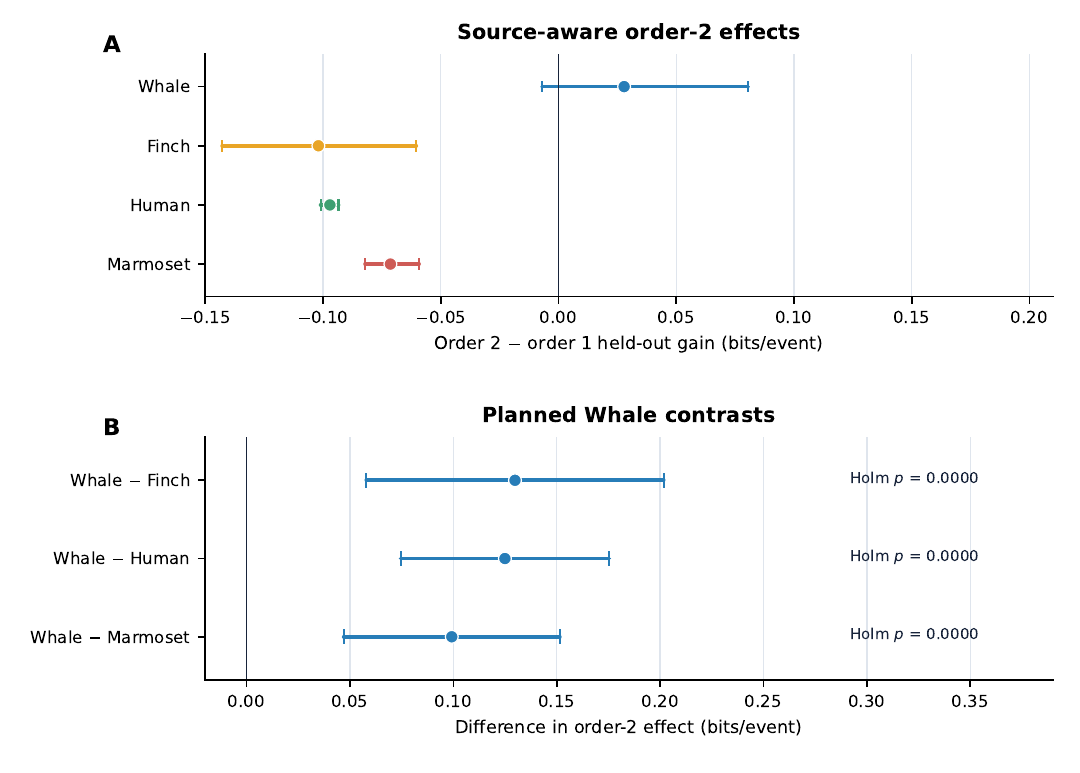}
\caption{Difference in held-out log likelihood for order-2 versus order-1 prediction on 1,307 common targets per draw, using training-source-only $K=32$ groups and family-balanced Ladder11 features. (A) Source-bootstrap 95\% intervals for each corpus; Whale's interval crosses zero. (B) Simultaneous 95\% intervals for the planned Whale-minus-reference contrasts; annotations show Holm-adjusted bootstrap tests conditional on the four Whale source pools. These tests are not equivalent to independent-animal replication. The exact two-sided four-pool sign-flip test cannot attain $p<0.125$.}
\label{fig:order}
\end{figure*}

Under the primary family-balanced Ladder11 representation, the order-2 minus order-1 held-out difference is $+0.0279$ bits/event for Whale (Fig.~\ref{fig:order}). Its source-bootstrap 95\% interval, $[-0.0069,+0.0805]$, includes zero; the exact two-sided four-pool sign-flip test gives $p=0.375$. We therefore cannot conclude that a second preceding event improves prediction for Whale alone. This distinction matters because four operational Whale source pools provide limited resolution for a source-level test.

Finch, Human, and Marmoset instead have negative changes of $-0.1019$, $-0.0971$, and $-0.0714$ bits/event. Consequently, the effect of using two predecessors is more favorable to Whale than to each reference. The planned Whale-minus-reference contrasts are $+0.1298$ for Finch, $+0.1249$ for Human, and $+0.0992$ bits/event for Marmoset. Their simultaneous 95\% bootstrap intervals are $[0.0579,0.2017]$, $[0.0745,0.1754]$, and $[0.0470,0.1515]$, respectively. These intervals are positive under the specified source-bootstrap procedure. With only four Whale source pools, however, the very small bootstrap tail probabilities must not be interpreted as strong independent-animal replication or as proof of a positive Whale-only effect.

The longer-history comparison also depends on how the acoustic groups are formed. Using the same capped-16 sequence blocks under fixed-seed fitting, the 23-view Literature23 panel gives $+0.1697$ bits/event for Whale, whereas the 11-configuration Ladder11 panel gives $+0.0197$. These controlled checks hold the sequence geometry fixed but change the representation and fitting setup. Separate fixed-tokenizer sensitivities are reported in the Supplementary Material. The primary four-corpus estimate remains $+0.0279$ bits/event, obtained with training-source-only group construction and family-balanced Ladder11. The sensitivity results help explain why the magnitude changes, but they do not replace the primary held-out estimate.

\subsection{Validation and control analyses}
The learned acoustic groups provide one description of the recordings. We next tested whether the patterns can also be seen in familiar waveform measurements, whether they persist without clustering, and how much they depend on source and event position. These checks are important because a common measurement procedure need not have the same sensitivity to every kind of recording.

\begin{figure*}[!t]
\centering
\includegraphics[width=0.97\textwidth]{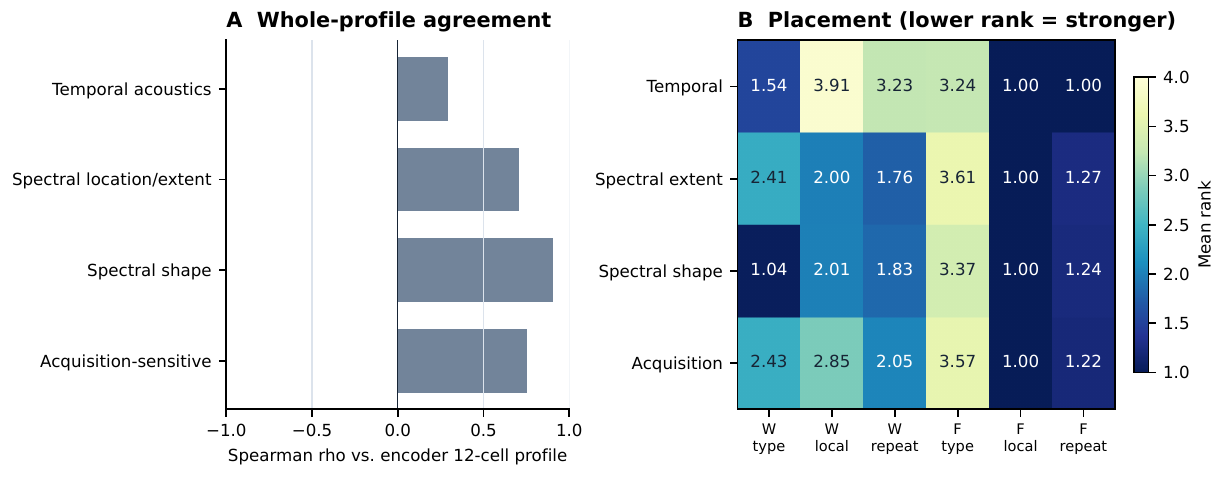}
\caption{Comparisons between the encoder profile and independently calculated waveform descriptors on the matched shell. (A) Spearman profile agreement across 12 dependent corpus-by-measurement cells; values are descriptive. (B) Mean ranks for Whale (W) and Finch (F) in type accumulation, local dependence, and repeated-order recurrence under each physical feature family. Smaller rank values indicate stronger placement. The physical and acquisition-sensitive families recover overlapping, but not identical, patterns.}
\label{fig:descriptors}
\end{figure*}

\medskip
\noindent\textit{Waveform measurements.}
Spectral shape gives the strongest agreement with the encoder-based profile, with descriptive Spearman $\rho=0.902$ across 12 dependent corpus-by-measurement cells (Fig.~\ref{fig:descriptors}). Spectral location and extent have $\rho=0.706$, while the temporal-acoustic family has $\rho=0.291$. These differences matter because each feature family tests a different aspect of the recorded sound. Spectral shape places Whale close to first for type accumulation, while temporal and spectral measurements recover strong Finch local and repeated order. The results show that familiar acoustic descriptions can capture parts of the profile without reproducing the same ordering under every measurement.

The temporal waveform comparison is particularly sensitive to the crop used for Whale. The primary annotation crops give a whole-profile Spearman correlation of $\rho=0.291$; widening the Whale windows and filling the added portions with zeros gives $\rho=0.469$. Using the actual recorded audio extending $\pm0.5$~s around the Whale events gives $\rho=0.762$. All three correlations compare the same 12 cells: four corpora crossed with type accumulation, local dependence, and recurrence. Only the Whale waveform measurements change; the reference-corpus measurements remain fixed. These are crop-context sensitivity results, not successive levels of biological validation.

However, acquisition-sensitive physical measures also agree with the encoder profile ($\rho=0.755$), and the combined nuisance-only family reaches $\rho=0.874$. Duration alone gives $\rho=0.455$ and padding-only measures give $\rho=0.434$. The correlations summarize a small, dependent set of 12 profile cells; they are not independent-species significance tests. They show that the learned representations contain properties of recorded sound, but that recording and duration-related features can reproduce substantial parts of the same profile. Agreement with waveform features therefore supports an acoustic measurement interpretation without isolating biological variation from acquisition effects. The difference between zero-filled and recorded-context windows also shows that surrounding audio changes the physical measurement. That context may include neighboring clicks, other sounds, or recording background; it cannot automatically be attributed to the annotated coda.

\medskip
\noindent\textit{Continuous acoustic analyses.}
Type accumulation relies on group labels, so we also measured how quickly the continuous physical-acoustic space is covered by additional observations. Larger coverage AUC indicates that sampled references cover the observed descriptor space more slowly. Without K-means, Whale has the largest pooled mean AUC (0.2994), followed by Human (0.2751), Marmoset (0.2262), and Finch (0.2168). Whale also has the largest source-residualized mean (0.4763), although the gap to Human (0.4573) is small. The result supports greater observed diversity for Whale in these measured acoustic descriptors, not a species-wide repertoire claim.

A corresponding continuous prediction test asks whether one event's acoustic properties help predict the next. Under the source-and-position null, Whale has excess held-out $R^2=0.2583$ and Finch 0.2327. Under the ordinary within-block null, Finch instead leads with 0.1387 compared with Whale 0.0420. The difference between these controls shows that there is no universal cross-corpus winner in continuous temporal prediction. The continuous analyses also use a physical descriptor space rather than encoder-generated discrete groups; they validate related properties of the recordings, not the identical categorical estimand.

\medskip
\noindent\textit{Sequence position and recording source.}
A group can appear predictable simply because it is concentrated in one recording or near one position in a song or block. We therefore compared ordinary randomization with source-and-position conditioning. Under that restricted null, the fixed-consensus neighboring excesses for Whale and Finch are statistically difficult to distinguish. The corresponding recurrence ratios depend on compressor: Finch leads RePair (0.777 versus Whale 0.811), while Whale has smaller LZ78 (0.833 versus 0.861) and DEFLATE (0.764 versus 0.943) ratios. The equal-weight restricted recurrence family favors Whale. Thus, the recurrence ordering depends on both the compressor and the null model.

Recording sources also affect acoustic accumulation. The Whale observations come from 63 recordings grouped into four conservative pools, including an unresolved remote-identity pool. They do not provide the same source granularity as the reference corpora. Removing source means changes continuous coverage but leaves measurable variation within recorded sources. Thus the comparison is best understood as a source-aware description of these recordings, not a controlled estimate of the biological repertoire of an entire species.

\medskip
\noindent\textit{Analysis choices.}
Switching from right-zero to symmetric-reflect padding preserves the broad Whale/Finch placement, and level-normalization checks show similar large-scale patterns. This reduces concern that zero padding alone produced the observed ordering. But both padding methods add artificial signal, and group assignments may carry information about duration even though the event sequence itself is unchanged. The results also vary in magnitude across clustering variants and encoder panels; sensitivity tables identify the accepted configurations rather than treating all analytical cells as independent samples. More detailed source-pool, representation, and fitting checks are reported in the Supplementary Material.
\section{Discussion}
The four corpora differ in which kinds of acoustic and temporal structure are most visible. The comparison is informative because the measurements are kept distinct and tested against recording- and source-sensitive alternatives.

\subsection{Different vocal systems reveal different forms of structure}
The clearest contrast is between the accumulation of acoustic groups and the recurrence of their sequences. Whale recordings show the fastest accumulation of encoder-defined acoustic groups under a matched event budget. Finch recordings show the strongest repeated-order recurrence against an ordinary block shuffle. These are different kinds of structure: one describes continued acoustic variation as observations are added, while the other describes how often group patterns repeat in the observed sequence.

The neighboring-event results make this distinction especially clear. When we shuffle order within blocks, Finch shows the larger excess dependence. When we also preserve source and relative position in the reference, the Whale and Finch estimates are closely matched. Restricted recurrence likewise changes depending on the compression procedure. A corpus can show strongly repeated sequences against an ordinary shuffle without leading every source-conditioned measure. These are not competing answers to one question. They are answers to different questions about how the recorded sounds relate over time.

The comparison is therefore most informative when the measurements are read together. Type accumulation describes the variety of encoder-defined acoustic forms. Local dependence describes relationships between adjacent groups. Recurrence describes the reuse of longer patterns. Neither strong acoustic diversity nor strong recurrence alone determines the other. The contrast between Whale and Finch demonstrates why these properties should be examined separately.

Human phones and marmoset calls also remind us that event scale matters. A phone is a small part of an utterance, whereas a coda, bird syllable, or marmoset call is defined on its own annotation scale. A weak recurrence measure for one kind of event says little about structure at a different scale. Our comparison is therefore about acoustic and sequential properties of these recorded units, not a hierarchy of communication systems.

\subsection{Frozen audio encoders can serve as common acoustic measurement tools}
A central methodological contribution is that frozen audio encoders can be used as common acoustic measurement tools. A pretrained encoder describes each event through many learned features, whether the event is a whale coda, a human phone, a bird syllable, or a marmoset call. The features exist before we form any acoustic groups within the study corpora. We can therefore examine how sounds vary in the encoder space, and then ask how those sounds are organized in time. The events retain their native definitions, and the same questions are asked of each corpus.

Pretrained audio models are often used to recognize sounds or to produce groups that can be checked against human annotations. Here the learned representation serves a different scientific purpose. We use it to measure acoustic diversity, neighboring-event relationships, and recurring sequences. Groups are introduced only for measurements that require categorical events; other analyses work directly with continuous acoustic properties. The particular group labels are intermediate parts of the measurement procedure, not claims about an animal's perceptual categories. This distinction makes the encoder useful as more than a way to separate sounds into clusters.

The direct waveform results help explain what the encoder-based measurements are capturing. Spectral-shape descriptors reproduce much of the overall profile, and the sequence relationships in Finch are also visible in physical acoustic measurements. Some differences remain in continuous analyses that do not require groups. These agreements connect the learned measurements to properties that can be observed in the recorded signals themselves.

The waveform comparisons also set important limits on that interpretation. Acquisition-sensitive and duration-related measurements recover substantial parts of the profile. The Whale crop-context test shows that widening the input window, even with zeros, can change the reported agreement. Adding genuine surrounding audio changes it further, but that audio may contain information beyond the annotated coda. The encoders are useful measurement tools because their outputs can be checked against these independent descriptions and controls, not because they are immune to the circumstances of recording.

Two kinds of comparability are therefore important. Matched event counts and sequence lengths give the corpora equal opportunities to exhibit the measured patterns. Applying the same encoder and statistical questions gives them a common analysis procedure. Neither step makes the acoustic distances, frequency ranges, or biological meanings of their events equivalent. The value of the framework is that it makes the differences measurable while keeping those boundaries visible.

\subsection{The structure that stands out depends on representation and temporal scale}
The amount of preceding context changes the temporal comparison. For A--B--C, the contrast between predicting C from B alone and predicting it from A and B is more favorable to Whale than to the other three corpora. The Whale-only estimate remains uncertain under its four source pools, so the result establishes a difference between corpora rather than an independently confirmed longer-history effect for Whale. This distinction prevents a relative result from becoming a claim about complex syntax or behavior.

The longer-history result also changes when we describe the sounds differently. The Literature23 representation gives a larger Whale estimate under its fixed fitting conditions than the family-balanced Ladder11 representation. Continuous next-event prediction changes the Whale--Finch ordering again when the reference shuffle is changed. These differences show that temporal organization is not one fixed property that an estimator either discovers or misses. A measurement can emphasize predictable transitions among discrete groups, predictable changes in continuous acoustic properties, or additional information from earlier events. We learn more by seeing where those answers agree and where they separate.

Together, the findings argue for examining several complementary properties rather than collapsing the recordings into a single ranking. Acoustic-group accumulation, repeated ordering, source-conditioned dependence, and continuous acoustic prediction each describe an observable aspect of vocal organization. Their differences are part of the scientific result, not merely technical variation to average away.
\section{Limitations}
The comparison uses four specific corpora at different native annotated event scales. Matched event counts and sequence opportunities make the stated statistical questions comparable, but cannot make codas, phones, syllables, and calls biologically equivalent. Neither the acoustic groups nor the resulting ranks should be interpreted as species-wide repertoire sizes or an ordering of communication-system complexity.

Recording conditions remain part of the measured variation. The Whale corpus mixes remote and tag recordings, and its 63 files are grouped into only four conservative source pools, including an unresolved-identity pool. These sources are not four verified whales or social units. The operational sequence blocks follow onset order within recordings, but available metadata do not verify complete continuous encounters or behavioral bouts. With only four Whale pools, the standalone longer-history source sign-flip test has limited resolution, and the bootstrap contrasts are conditional on a small number of operational source units.

Short Human and Finch events often need padding to reach the 0.175-s encoder input window. A separate input-exposure audit of the prior sampling protocol found that 93.1\% of Human events and 98.9\% of Finch events needed padding, comprising 51.7\% and 49.2\% of their encoder inputs. These percentages are not estimates for the present matched sample. Padding leaves the event order unchanged but can change group assignments and emphasize duration-related patterns. The broad Whale/Finch comparison persists with reflected input, and independent spectral waveform measurements recover strong Finch order. These checks weaken a zero-padding-only explanation, but do not establish freedom from short-event, duration, or acquisition effects.

The encoders were pretrained on different audio domains, including human speech, and their exact checkpoint histories still require verification. Even a frozen model can have domain-specific sensitivity. The physical spectral comparison uses a common 0--8-kHz band and therefore does not describe all frequency content of the native recordings. Widening Whale waveform crops changes temporal-profile agreement even when the added region is filled with zeros. Real recorded context changes it further, but can include neighboring vocal events and background sound. These sensitivities do not establish where the measured temporal information originates. Further source-resolved recordings, verified checkpoint manifests, and comparisons across event scales would strengthen biological and methodological generality.
\section{Conclusion}
The matched comparison shows that acoustic and temporal structure appear differently in four vocal corpora. Whale recordings have the fastest accumulation of encoder-defined acoustic groups, while Finch recordings have the strongest repeated-sequence recurrence against an ordinary block shuffle. Neighboring-event dependence becomes closely matched under a source-and-position control, and an additional preceding event changes held-out prediction more favorably for Whale than for the reference corpora. The Whale-only longer-history effect remains uncertain.

More broadly, pretrained audio encoders provide a way to measure acoustic variation and event order across very different recordings. Using several measurements reveals distinctions that a single ranking would miss. Matched sampling, waveform checks, and source-aware controls show how far those comparisons can be taken. Together, they offer a practical framework for studying how different forms of vocal structure appear in recorded sound.
\section*{Supplementary material}
The supplementary material reports the matched-shell reconstruction, estimator and inference definitions, complete source-pool audit, waveform and continuous analyses, position-conditioned results, input-length and encoder sensitivities, and an indexed map of all 27 replacement numerical tables. Machine-readable tables and analysis files accompany the source archive.

\begin{acknowledgments}
We thank the maintainers of the source recordings and annotations and the developers of the pretrained audio encoders. This research received no external funding.
\end{acknowledgments}

\section*{Author declarations}
\textbf{Conflict of interest.} The authors declare no conflicts of interest. 

\textbf{Ethics approval.} This study is a secondary analysis of existing recordings and metadata. Ethics and collection permissions for the original recordings are described by their source distributors and publications; the present study did not collect new animal or human recordings.

\textbf{Use of AI-assisted tools.} AI-assisted language tools were used for editorial drafting and formatting. The authors are responsible for verifying the scientific content, provenance, citations, analyses and final manuscript before submission.

\section*{Data availability}
The Whale audio and annotations are linked in the public SpermWhales collection~\citep{gubnitskyDiamantPublicSpermWhales}. The accompanying manifest identifies all 63 recording--annotation pairs by filename, public file ID, byte size, and acquisition-snapshot SHA-256 checksum. A dataset-specific DOI and redistribution rights have not been verified. The research archive provides the matched-shell audit, numerical tables, scripts, and checksums; native audio and pretrained weights are not redistributed. Regenerating embeddings requires the original files and exact encoder checkpoints.

\bibliographystyle{JASAauthyear2}
\bibliography{references}

\end{document}